\documentstyle[twocolumn]{article}
\makeatletter
\def\@normalsize{\@setsize\normalsize{12pt}\xpt\@xpt
\abovedisplayskip 10pt plus2pt minus5pt\belowdisplayskip \abovedisplayskip
\abovedisplayshortskip \z@ plus3pt\belowdisplayshortskip 6pt plus3pt
minus3pt\let\@listi\@listI} 

\def\subsize{\@setsize\subsize{12pt}\xipt\@xipt}

\def\section{\@startsection {section}{1}{\z@}{24pt plus 2pt minus 2pt}
{12pt plus 2pt minus 2pt}{\large\bf}}

\def\subsection{\@startsection {subsection}{2}{\z@}{12pt plus 2pt minus 2pt}
{12pt plus 2pt minus 2pt}{\subsize\bf}}
\makeatother

\begin{document}

\date{}

\title{\Large\bf A Review of Different Word Embeddings for Sentiment Classification using Deep Learning}

 
\author{\begin{tabular}[t]{c@{\extracolsep{8em}}c}
  \\     \hspace{3cm}Debadri Dutta \\
 \\
        \hspace{3cm}School of Electronics \\
        \hspace{3cm}Kalinga Institute of Industrial Technology \\
        \hspace{3cm} Bhubaneswar-751024, Odisha, India
\end{tabular}}

\maketitle

\thispagestyle{empty}

\subsection*{\centering Abstract}
{\em
The web is loaded with textual content, and Natural Language Processing is a standout amongst the most vital fields in Machine Learning. But when data is huge simple Machine Learning algorithms are not able to handle it and that is when Deep Learning comes into play which based on Neural Networks. However since neural networks cannot process raw text, we have to change over them through some diverse strategies of word embedding. This paper demonstrates those distinctive word embedding strategies implemented on an Amazon Review Dataset, which has two sentiments to be classified: Happy and Unhappy based on numerous customer reviews. Moreover we demonstrate the distinction in accuracy with a discourse about which word embedding to apply when.
\\Keywords: Natural Language Processing, Machine Learning, Word Embedding, Neural Networks
}

\section{Introduction}

Semantic vector space models of language represent
each word with a real-valued vector. These vectors can be utilized as highlights in multiple applications, for example, data recovery
document classification, sentiment classification, parsing, text generation, etc.

Word embeddings are in certainty a class of methods where singular words are represented to as real-valued vectors in a predefined vector space. Each word is mapped to one vector and the vector values are found out in a way that takes after a neural network, and subsequently the procedure is frequently lumped into the field of deep learning. Key to the approach is utilizing a dense distributed representation for each word. 
Each word is represented to by a real-valued vector, often tens or many measurements. This is differentiated to the thousands or millions of dimensions required for sparse word representations, for example, a one-hot encoding.

The popular models that we are aware of are, the skipgram method, the CBOW model under the word2vec, the GloVe embedding method.
In this work we analyze the different word embedding models,  on an Amazon Review Dataset, for our deep learning model, and display the results obtained in the accuracy levels.
\section{An Overview of the Different Word Embeddings}
 \hspace{-0.6cm} \textbf{Embedding Layer:} An embedding layer, for absence of a superior name, is a word embedding that is found out mutually with a neural network show on a particular natural language processing task, for example, language modelling or document classification. It requires that document text be cleaned and arranged with the end goal that each word is one-hot encoded. The span of the vector space is indicated as a component of the model, for example, 50, 100, or 300 measurements. The vectors are introduced with small random numbers. The embedding layer is utilized toward the front of a neural network and is fit supervisedly utilizing the Backpropagation calculation. The one-hot encoded words are mapped to the word vectors.  In the case if a recurrent neural network is utilized, at that point each word might be taken as one input in a sequence.This approach of learning an embedding layer requires a lot of training data and can be slow, but will learn an embedding both targeted to the specific text data and the NLP task.
 
 \hspace{-0.6cm} \textbf{GloVe Embedding:}The Global Vectors for Word Representation, or GloVe, calculation is an augmentation to the word2vec strategy for effectively learning word vectors, created by Pennington, et al. at Stanford. Classical  vector space models portrayals of words were produced utilizing matrix factorization strategies, for example, Latent Semantic Analysis (LSA) that complete a great job of utilizing global text statistics yet are not in the same class as the educated techniques like word2vec at catching importance and exhibiting it on undertakings like figuring analogies. 
GloVe is an approach to marry both the worldwide measurements of matrix factorization procedures like LSA with the  local context-based learning in word2vec.
As opposed to utilizing a window to characterize nearby setting, GloVe builds an express word-context or word co-occurence matrix utilizing statistics over the entire text corpus. The outcome is a learning model that may bring about for the most part better word embeddings.\vspace{1mm}
 
\hspace{-0.6cm} \textbf{Word2Vec:} Word2Vec is a statistical method for efficiently learning a standalone word embedding from a text corpus. In the year 2013, Tomas Mikolov, et al. whle working in Google came up with a solution to make embedding training more efficient with pre-trained word embedding. It deals with mainly two processes: 

i) \hspace{0.2cm}Continuous  Skip-Gram Model

ii) \hspace{0.15cm}Continuous Bag-of-Words Model or CBOW 

\hspace{-0.45cm}However here, in this case I have worked only with the CBOW Model.

\hspace{-0.6cm} \textbf{Difference b/w word2vec \& GloVe Embedding:}

The essential distinction amongst word2vec and GloVe embedding is that, word2vec is a "predictive" model though GloVe embedding is a "count-based" model. Predictive models take in their vectors so as to enhance their predictive capacity of Loss(target word | setting words; Vectors), i.e. the loss of predicting the target words from the context words given the vector representations. In word2vec, this is given a role as a feed-forward neural system and streamlined all things considered utilizing SGD, and so on. 

Count-based models take in their vectors by basically doing dimensionality reduction on the co-occurrence counts matrix. They first build an extensive network of (words x context) co-occurrence information, i.e. for each "word" (the lines), you count how as often as possible we see this word in some "specific circumstance" (the columns) in a vast corpus. The number of "contexts" is obviously extensive, since it is basically combinatorial in estimate. So then they factorize this matrix to yield a lower-dimensional (word x highlights) matrix, where each row currently yields a vector representation for each word. All in all, this is finished by limiting a "reconstruction loss" which attempts to discover the lower-dimensional representations which can clarify the greater part of the variance in the high-dimensional information. In the particular instance of GloVe, the count matrix is preprocessed by normalizing the counts and log-smoothing them. This ends up being a good thing as far as the  quality of the learned representations.

\section{Results and Conclusions}

\hspace{-0.5cm}The methods were implemented on an Amazon Review Dataset,which had almost 1 million words and 0.72 million sentences posted by the Customers. There were two sentiments to be classified: Happy and Unhappy. For each method, the dataset was divided into 70\% train data and 30\% test data and the training was done with only 2 epochs on CPU. However, for each case it took almost 3-4 hours on an average for each epoch to complete.

 \hspace{-0.6cm} \textbf{Embedding without pre-trained weights:}
 
 \hspace{-0.5cm}The output vectors are not processed from the input information utilizing any mathematical function. Rather, each information number is utilized as the index to get to a table that contains every possible vector. That is the motivation behind why you have to indicate the size of the vocabulary as the primary contention.

 \hspace{-0.6cm} \begin{tabular}{ |p{2cm}|p{2cm}|p{2cm}| }
 \hline
 \multicolumn{3}{|c|}{Embedding w/o pre-trained weights} \\
 \hline
 Epoch No. & Accuracy(\%) &Validation Accuracy(\%)\\
 \hline
1   & 94.33    &94.64\\
 2&   97.60  & 95.40\\
 \hline
\end{tabular}\vspace{5mm}

 \hspace{-0.6cm} \textbf{GloVe Embedding:}
 
 \hspace{-0.5cm}The insights of word events in a corpus is 
the essential wellspring of data accessible to all unsupervised techniques for learning word representations, furthermore, albeit numerous such techniques presently exist, the inquiry still stays with respect to how meaning is produced from these measurements, and how the subsequent word vectors may speak to that significance. We utilize our bits of knowledge to develop another model for word portrayal which we call GloVe, for Global Vectors, in light of the fact that the global corpus insights are caught straightforwardly by the model.

 \hspace{-0.6cm} \begin{tabular}{ |p{2cm}|p{2cm}|p{2cm}|  }
 \hline
 \multicolumn{3}{|c|}{GloVe Embedding} \\
 \hline
 Epoch No. & Accuracy(\%) &Validation Accuracy(\%)\\
 \hline
1   & 82.07    &79.91\\
 2&   85.20  & 83.32\\
 
 \hline
\end{tabular}\vspace{5mm}

\hspace{-0.6cm} \textbf{Embedding with Word2Vec CBOW \& Negative Sampling:}
 
 \hspace{-0.5cm}The goal of word2vec is to discover word embeddings, given a text corpus. As it were, this is a strategy for discovering low-dimensional representations of words. As an outcome, when we discuss word2vec we are regularly discussing Natural Language Processing (NLP) applications. For instance, a word2vec demonstrate prepared with a 3-dimensional hidden layer will bring about 3-dimensional word embeddings. It implies that, say, "apartment" will be represented by a three-dimensional vector of real numbers that will be close (consider it regarding Euclidean separation) to a comparable word, for example, "house". Put another way, word2vec is a procedure for mapping words to numbers. There are two fundamental models that are utilized inside the setting of word2vec: the Continuous Bag-of-Words (CBOW) and the Skip-gram show. Here the experiment was done only with the CBOW model along with negative sampling. In the CBOW model the objective is to find a target word, given a context of words. In the simplest case in which the word’s context is only represented by a single word. 

 \hspace{-0.6cm} \begin{tabular}{ |p{2cm}|p{2cm}|p{2cm}|  }
 \hline
 \multicolumn{3}{|c|}{Embedding with Word2Vec 
 
 CBOW \& Negative Sampling} \\
 \hline
 Epoch No. & Accuracy(\%) &Validation Accuracy(\%)\\
 \hline
1   & 80.33    &82.98\\
 2&   85.88  & 86.53\\
 
 \hline
\end{tabular}\vspace{5mm}

\subsection*{Conclusion}
The astonishing actuality was that Embedding with no pre-trained weights had a superior outcome than word2vec with pre-trained weight or GloVe Embedding. This is a territory where additionally tests can be done, most likely an a whole lot greater dataset or for different purposes like text generation. In any case, for sentiment classification in light of Customer surveys, pre-trained weights couldn't satisfy that desires, which can be comprehended by implies for some examination.


\begin{thebibliography}{9}

\bibitem{key:foo}
``Global Vectors for word Representation''
https://nlp.stanford.edu/pubs/glove.pdf

\bibitem{key:foo}
``Linguistic Regularities in Continuous Space Word Representations''
https://www.microsoft.com/en-us/research/wp-content/uploads/2016/02/rvecs.pdf

\bibitem{foo:baz}
``Efficient Estimation of Word Representations in
Vector Space''
https://arxiv.org/pdf/1301.3781.pdf
\bibitem{foo:baz}
``Distributed Representations of Words and Phrases
and their Compositionality''
https://arxiv.org/pdf/1310.4546.pdf
\bibitem{foo:baz}
``Neural Network Methods in Natural Language Processing (Synthesis Lectures on Human Language Technologies) by Yoav Goldberg'' 
\bibitem{foo:baz}
``A systematic comparison of
context-counting vs. context-predicting semantic vectors''
http://clic.cimec.unitn.it/marco/publications/acl2014/baroni-etal-countpredict-acl2014.pdf
\end{thebibliography}
\end{document}